# Electrically Reconfigurable Dual Metal-Gate Planar Field-Effect Transistor for Dopant-free CMOS

Tillmann Krauss, Frank Wessely and Udo Schwalke
Institute for Semiconductor Technology and Nanoelectronics
Technische Universität Darmstadt
Darmstadt, Germany
krauss@iht.tu-darmstadt.de

*Abstract*— In this paper, we demonstrate by simulation the feasibility of electrostatically doped and therefore reconfigurable planar field-effect-transistor (FET) structure which is based on our already fabricated and published Si-nanowire (SiNW) devices. The technological cornerstones for this dual-gated general purpose FET contain Schottky S/D junctions on a silicon-on-insulator (SOI) substrate. The transistor type, i.e. n-type or p-type FET, is electrically selectable on the fly by applying an appropriate control-gate voltage which significantly increases the versatility and flexibility in the design of digital integrated circuits.

## I. Introduction

The dominating leakage path in today's scaled MOSFET devices for OFF-state currents stem from PN-junction and bulk-leakage [1], [2]. Both currents increase with temperature. SOI based technologies can reduce bulk-leakage currents significantly but junction leakage still remains present even in SOI FETs. Therefore, our device concept includes no PN-junctions but Schottky-barrier source/drain contacts adding to the high temperature robustness as shown previously in refs. [3], [4] and [5].

The intrinsic combination of an ambipolar device behavior with a separated current flow control-gate results in a remarkable leakage current suppression and enhanced on-to-off current ratio. Additionally, the flexibility to instantly select n-channel and p-channel behavior opens a new way for designing reconfigurable integrated circuits [10]. Furthermore, as no standard CMOS doping process is necessary, the device does not deteriorate due to statistic dopant fluctuation and dopant dependent reduction of carrier mobility that typically arises in up-to-date aggressively scaled MOSFET devices.

To the best of our knowledge, all ambipolar devices with a separate control electrode to regulate the current flow involve nanowire structures as, for example, silicon nanowires or carbon nanotubes. A summary of reconfigurable silicon electron devices is given in reference [6]. We want to emphasize that the discussed planar device concept in this paper is quite different.

## II. Device Structure

The crosscut of the designed device structure is shown in Fig. 1. It is partly based on our published SiNW FET technology [3], [4], [5]. The new concept, as well as the previous the SiNW technology, is experimentally based on a virtually undoped SOI substrate, i.e. with the lowest commercially available background boron doping of $2 \times 10^{15}$ cm$^{-3}$ [7].

The simulated device structure is built on a SOI substrate with a 40 nm top silicon and 50 nm buried silicon oxide layer (BOX). It possesses three gate electrodes, namely front-gate (FG) M1, FG M2 and back-gate (BG). Note, that FG M1 and FG M2 are electrically shorted but can differ in work function. The handle silicon layer (support wafer) functions as the BG contact and is insulated by the BOX from the top silicon layer. The FG M1 and M2 are positioned into a recess in the center of the top silicon with 100 nm gate length each or 200 nm total ($L_{FG}$ in Fig. 1). The FG is insulated by a hafnium oxide layer of 10 nm thickness and the channel height is 10 nm if not mentioned otherwise. Finally, the source and drain contacts are placed and connected to the top silicon layer with a lateral distance of 1 µm ($L_{BG}$ in Fig. 1).

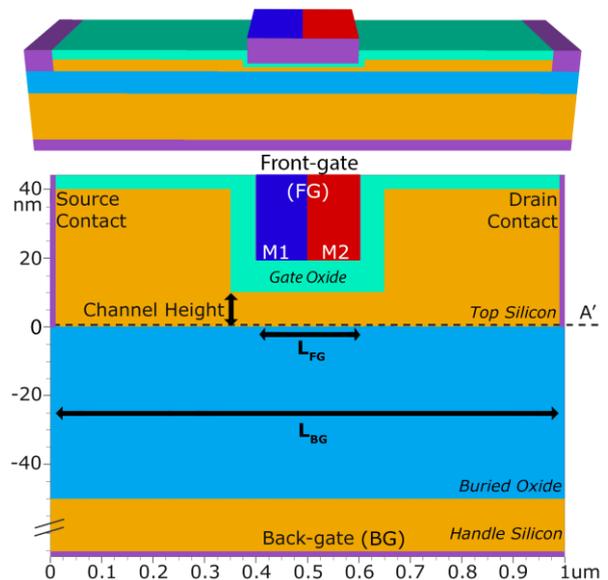

Fig. 1. Schematic cross-section of the simulated planar dual metal front-gate FET structure.

## III. RESULTS AND DISCUSSION

In this chapter, we will demonstrate by simulation the various effects of biasing conditions, physical geometry variations as well as FG metal selection (i.e. work-function) on the electrical characteristics of the device. The 2D simulations are carried out using Silvaco Atlas and its drift-diffusion models.

To facilitate the understanding of the device operation, the FET is regarded as an intrinsic combination of two interacting transistors. Each of these two transistors is represented by its own gate electrode, the FG (M1 and M2) and BG respectively.

The transistor formed by the BG electrode influences the whole device channel region as an ambipolar enhancement mode (normally-off) transistor over the full back-gate length ($L_{BG}$ in Fig. 1). This will define the transitor type, i.e. NMOS or PMOS. On the contrary, the influence of the second transistor formed by the FG electrodes M1 and M2 working in depletion mode (normally-on) is limited to the center of the channel region ($L_{FG}$ in Fig. 1) and will affect the current flow between S/D. The combination of the two operation modes can be termed as dehancement mode operation.

### A. Back-Gate Enhancement Mode Operation

Figure 2 depicts a simulated BG voltage sweep with an electrically floating FG and a constant source bias of $V_S = 0.1$ V displaying an ambipolar characteristic. Applying a sufficiently high BG bias voltage ($V_{BG}$) results in an increasing source current as either holes ($V_{BG} < 0.5$ V) or electrons ($V_{BG} > 1$ V) are attracted to the top-silicon-to-BOX interface forming a channel.

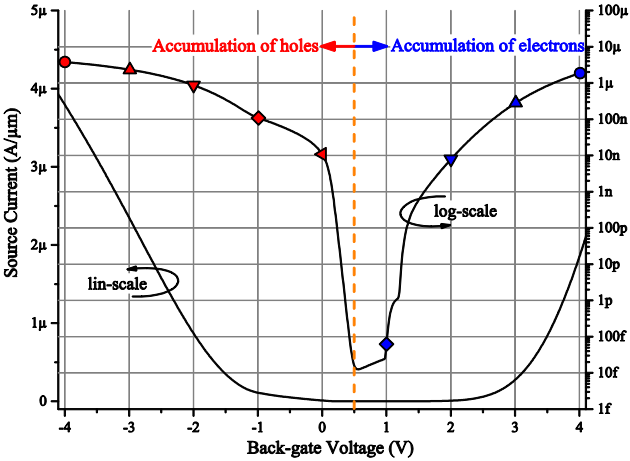

Fig. 2. Voltage sweep of the back-gate contact with floating front-gate condition at a source voltage of $V_S = 0.1$ V. Symbol shapes correspond to operating points in Fig. 6.

Figure 3 and 4 depict two BG biasing conditions as band diagrams for a single front-gate work function of 4.7 eV. For a negative BG voltage, holes are attracted and accumulating at the BOX interface (Fig. 3). Vice versa, an accumulation of electrons takes place for a sufficiently positive BG voltage (Fig. 4). This resembles a p- respectively n-type enhancement mode or accumulation mode transistor.

The charge carriers originate from the mid-gap Schottky barrier S/D contacts whose work functions are set to 4.6 eV. The Schottky barrier shape and therefore probability for tunneling and thermionic field emission of electrons respectively holes is influenced by the electrical field of the BG. A higher magnitude of BG voltage reduces the Schottky barrier width and increases the tunneling current through the barrier.

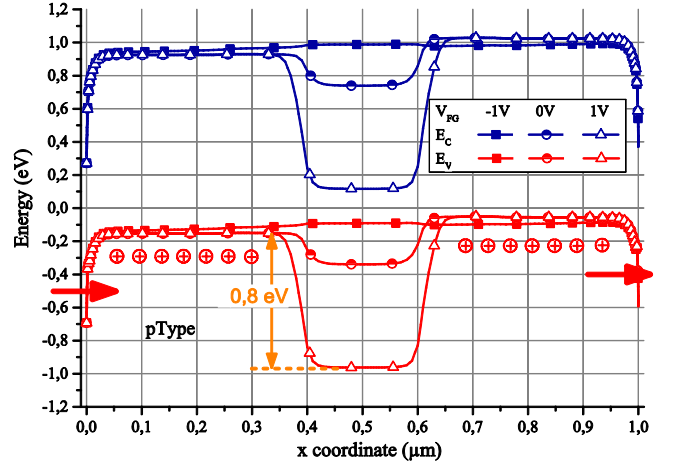

Fig. 3. Band diagram for a negative back-gate biasing $V_{BG} = -3$ V at a source voltage of $V_S = 0.1$ V for a single metal front-gate (M1 = M2) with a work function of 4.7 eV. Filled symbols show on-state transistor, partly filled symbols represent transition state. Open symbols visualise off-state.

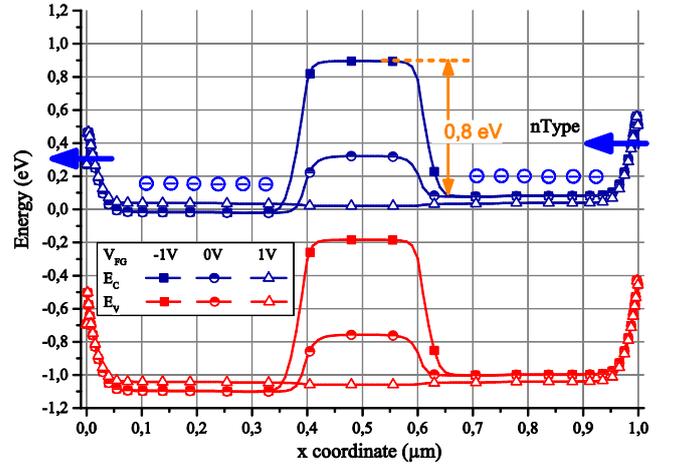

Fig. 4. Band diagram for a positive back-gate biasing $V_{BG} = +3$ V at a source voltage of $V_S = 0.1$ V for a single metal front-gate (M1 = M2) with a work function of 4.7 eV. Open symbols show on-state transistor, partly filled symbols represent transition state. Filled symbols visualise off-state.

The Schottky barrier parameters for the simulation of the S/D contacts have been estimated by comparing and fitting simulated and measured back-gate voltage sweeps of the fabricated SiNW FET devices with 70 nm x 70 nm rectangular diameter and 50um length (Fig. 5) [3], [4], [5]. Remaining deviations between measurement and simulation are mainly caused by differences in device and contact geometries as well as not modeled complex Schottky barrier effects (not shown here) as, for example, barrier lowering due to image charges.

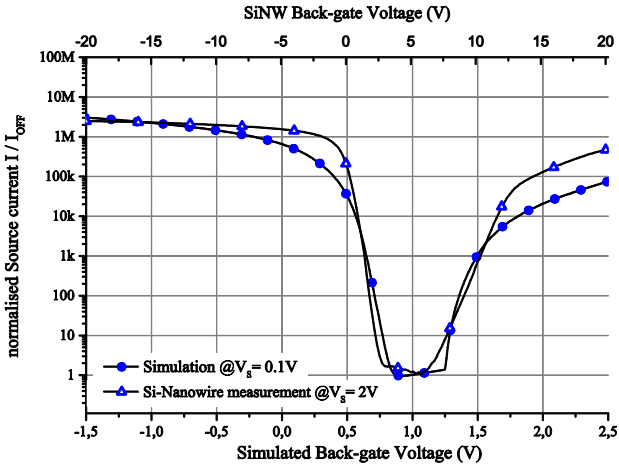

Fig. 5. Comparison of back-gate voltage sweeps of simulation and measurement data of silicon nanowire FET both with electrically floating front-gate condition.

## B. Front-Gate Depletion Mode Operation

The FG controls the flow of charge carriers through the accumulated (i.e. electrostatically doped) channel formed by the BG. Therefore, the FG transistor resembles a depletion mode FET.

The current-voltage characteristics for FG voltage ($V_{FG}$) sweeps for different BG biasing conditions are shown in Fig. 6. The simulation shows an on-to-off current ratio of up to ~11 decades with a typical leakage current close to 1 aA/μm. These low leakage currents are the result of high potential barriers for electrons and holes of up to 0.8 eV formed by the FG electrical field depicted in the band diagrams in Fig. 3 and Fig. 4. The maximum on-current on the other hand is directly correlated to the magnitude of back-gate voltage.

Furthermore, the threshold voltage can be shifted by approximately 100 mV/V back-gate voltage granting circuit designers additional design flexibility (Fig. 6).

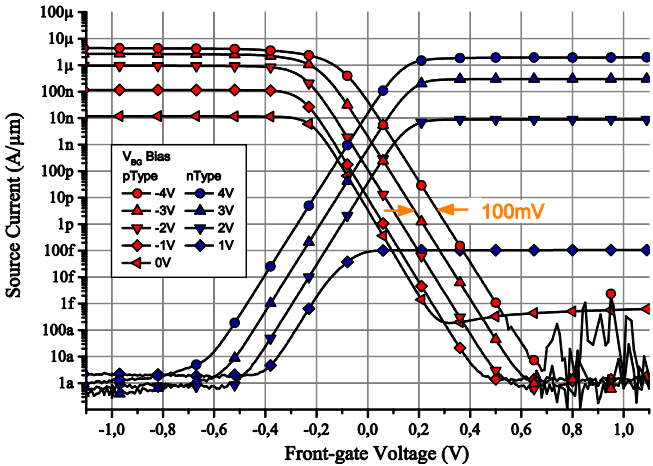

Fig. 6. Voltage sweeps of single metal front-gate with a work function of 4.7 eV at $V_S = 0.1$ V for different back-gate voltages. Symbol shapes correspond to operating points in Fig. 2.

## C. Effect of Channel Height and Front-Gate Length

The electrical effect of the FG is primarily determined by the distance, length and effective electric permittivity of the material stack between the BG channel and the FG electrode.

The off-state leakage current is strongly correlated to the channel height. An increase of the channel height from 10 to 30 nm significantly degrades leakage suppression from ~8 aA/μm to ~2 nA/μm reducing the on-to-off current ratio for p- and n-type operation from ~12 down to ~3 decades as depicted in Fig. 7. For this reason, a homogeneous recess fabrication with a defined channel height in the front-gate region is a key process step in manufacturing of the proposed device.

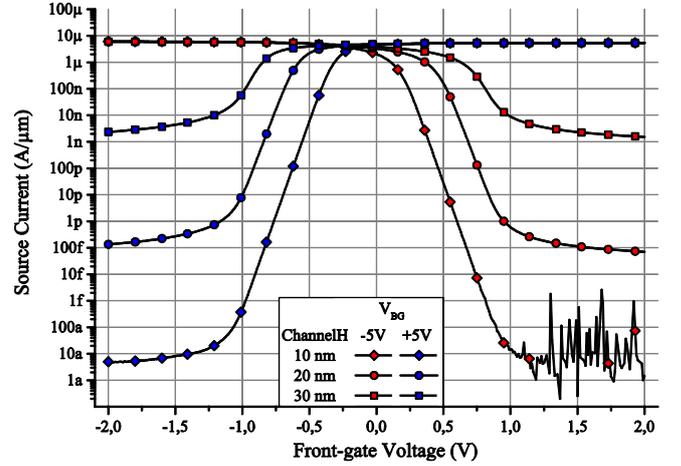

Fig. 7. Effect of channel height on leakage supression performance for $V_S = 0.1$ V for a 10 nm silicon nitride front-gate insulation.

As expected, a reduction of the FG length ($L_{FG}$) considerably degrades the subthreshold slope and increases the threshold voltage. However, when using a high-k oxide like hafnium oxide instead of silicon oxide one can mitigate the degradation as depicted in Fig. 8 for $L_{FG}$ 10 and 20 nm with hafnium oxide (open symbols).

As the channel has no direct interface with the FG oxide and no high temperature processing steps are required after FG oxide formation there are no technological issues like interface deterioration and temperature instability of high-k insulators which is known to degrade the channel when high-k oxides like gadolinium oxide or other lanthanide oxides are introduced as gate insulators [8], [9] for example.

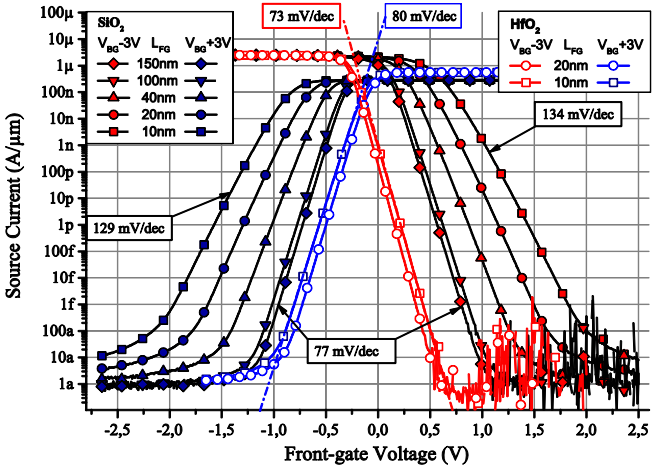

Fig. 8. Effect of front-gate length and permittivity oxide on subthreshold slope and theshold voltage for a source voltage of $V_S = 0.1$ V and a single metal front-gate with a work function of 4.6 eV (M1 = M2).

## D. Effect of Dual Metal Front-Gate

Another parameter influencing the electrical effect of the FG is the effective work function of the FG metal. The metal work function is directly linked to the threshold voltage characteristic of field-effect transistors.

The influence of different work functions for a single metal FG with $L_{FG}$ 200 nm is depicted in Fig. 9. A work function difference of 0.6 eV results in an simultaneous one-to-one shift of the threshold voltage of 0.6 V for p- (red symbols) and n-type (blue symbols) operation mode.

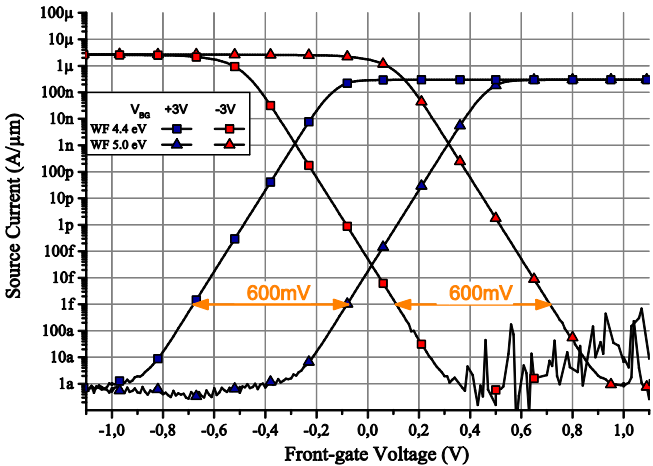

Fig. 9. Effect of single metal front-gate ($L_{FG}$ 200 nm) on threshold voltage with metal work funcitons of 4.4 eV and 5.0 eV.

In order to reduce switching currents in CMOS designs it is mandatory to technologically shift and control the threshold voltages separately for n- and p-type transistors. A combination of two metals with different work functions in a dual metal FG can satisfy this requirement as shown in Fig. 10.

Using a dual metal FG can reduce the crossing current for a FG voltage of 20 mV from 1 nA/μm for 4.7 eV single metal FG down to 100 fA/μm for a dual metal FG with work function of 4.4 eV for M1 and 5 eV for M2 (Fig. 10).

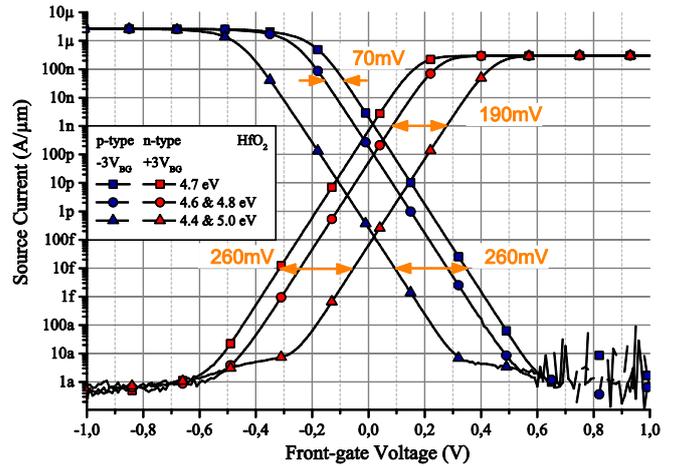

Fig. 10. Effect of dual metal front-gate ($L_{FG}$ 2x100 nm) metal work function difference on threshold voltage.

Because of the reduced effective FG length from 1x 200 nm to 2x 100 nm ($L_{FG}$ M1 = $L_{FG}$ M2 = 100 nm) the threshold voltage shift is not a one-to-one relation to metal work function differences. The threshold voltage shift can also be derived from the band diagrams depicted in Fig. 11 and Fig. 12. For a FG voltage of 0 V (half-filled symbols) only the potential barrier of either M1 or M2 is interfering with the charge carrier flow through the channel.

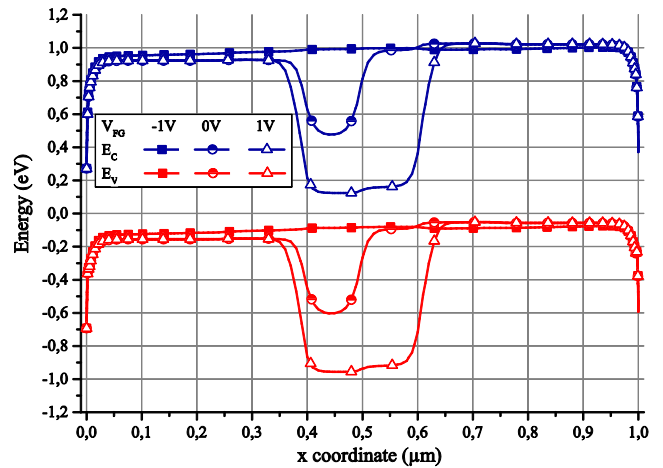

Fig. 11. Band diagram for negative back-gate biasing $V_{BG}$ = -3 V at a source voltage of $V_S$ = 0.1 V for dual metal front-gate with metal work functions M1 = 4.4 eV and M2 = 5.0 eV. Filled symbols show on-state transistor, half-filled symbols represent transition state. Open symbols visualise off-state.

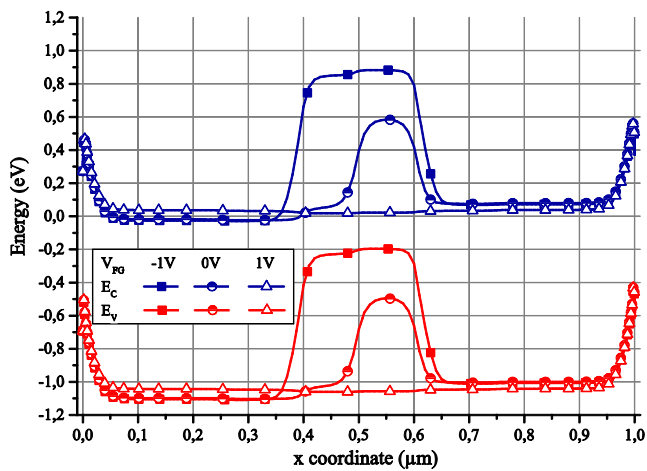

Fig. 12. Band diagram for positive back-gate biasing $V_{BG} = +3$ V at a source voltage of $V_S = 0.1$ V for a dual metal front-gate with work functions M1 = 4.4 eV and M2 = 5.0 eV. Open symbols show on-state transistor, half-filled symbols represent transition state. Filled symbols visualise off-state.

## IV. Conclusion

In conclusion, the simulated device with midgap Schottky-barrier source and drain contacts and dual metal front-gate can serve as a versatile supplemental building block for logic and analog designs. The dual metal front-gate approach possess a remarkable off-state leakage current in combination with an expected reduction of threshold voltage variances due to the absence of doping fluctuations which is a stepping-stone for ultra low power silicon CMOS designs. The capability to select electrically the transistor type (i.e. PFET or NFET) on the fly, enables digital circuit designers to build switchable NAND/NOR cells with the very same physical transistors in place saving precious silicon area [10]. Using innovative logic synthesis methodologies like MIXSyn the total transistor count for logic applications can be further reduced [11], [12].